\documentclass[submission,copyright,creativecommons]{eptcs}
\providecommand{\event}{SOS 2007} %

\usepackage{iftex}

\ifpdf
  \usepackage{underscore}         %
  \usepackage[T1]{fontenc}        %
\else
  \usepackage{breakurl}           %
\fi

\title{Minimal Model Counting via Knowledge Compilation}
\author{Mohimenul Kabir
\institute{National University of Singapore}
\institute{School of Computing\\
Singapore}
}
\def\titlerunning{Minimal Model Counting via Knowledge Compilation}
\def\authorrunning{Kabir}
\usepackage{amsmath}
\usepackage{amsfonts}
\usepackage{cleveref}
\usepackage{flexisym}
\usepackage{xfrac}
\usepackage{algorithm}
\usepackage{algpseudocode}
\usepackage{bbm}
\usepackage{xspace}

\newcommand{\sharpASP}{\ensuremath{\mathsf{sharpASP}}}
\newcommand{\forced}[1]{\mathsf{Forced}(#1)}
\newcommand{\copyoperation}[1]{\mathsf{Copy}(#1)}
\newcommand{\copyatom}[1]{#1\textprime}
\newcommand{\loopatoms}[1]{\mathsf{LA}}
\newcommand{\dependency}[1]{\mathsf{DG}(#1)}
\newcommand{\mmodel}[1]{\mathsf{MinModels}(#1)}
\newcommand{\Card}[1]{|#1|}
\newcommand{\Sol}[1]{\mathsf{Sol}(#1)}
\newcommand{\be}{\mathbb{B}}
\newcommand{\conpcomplete}{$\mathsf{co}$-$\mathsf{NP}$-complete}
\newcommand{\cntmm}[1]{\mathsf{CntMM}(#1)}
\newcommand{\true}{\ensuremath{\mathsf{true}}\xspace}
\newcommand{\false}{\ensuremath{\mathsf{false}}\xspace}

\newcommand{\clause}[1]{\mathsf{Clause}(#1)}
\newcommand{\var}[1]{\mathsf{Var}(#1)}
\newcommand{\toolname}{\ensuremath{\mathsf{KC}}-\ensuremath{\mathsf{min}}}

\algnewcommand\algorithmicforeach{\textbf{for each}}
\algdef{S}[FOR]{ForEach}[1]{\algorithmicforeach\ #1\ \algorithmicdo}
\newcommand{\np}{\mathsf{NP}}
\newcommand{\co}{\mathsf{co}}
\newcommand{\p}{\mathsf{P}}
\newtheorem{lemma}{Lemma}
\newtheorem{definition}{Definition}
\newtheorem{proposition}{Proposition}
\newtheorem{proof}{Proof}
\newtheorem{example}{Example}

\begin{document}
\maketitle
\begin{abstract}
  Counting the number of models of a Boolean formula is a fundamental problem in artificial intelligence and reasoning. 
Minimal models of a Boolean formula are critical in various reasoning systems, making the counting of minimal models essential for detailed inference tasks. 
Existing research primarily focused on decision problems related to minimal models. 
In this work, we extend beyond decision problems to address the challenge of counting minimal models. 
Specifically, we propose a novel knowledge compilation form that facilitates the efficient counting of minimal models. 
Our approach leverages the idea of justification and incorporates theories from answer set counting.

\end{abstract}
\section{Introduction}
Given a Boolean formula, the minimal model refers to its subset minimal models~\cite{Mccarthy1980}.
The minimal model has applications in fundamental reasoning tasks, including circumscription~\cite{Lifschitz1985}, diagnosis~\cite{DMR1992}, deductive database~\cite{Minker1982}, and like. 
While existing literature primarily addresses decision-version problems related to minimal models~\cite{ABFL2017}, our work focuses on the counting-version of the minimal model problem~\cite{KM2024}. 
This counting problem provides a more fine-grained mode of reasoning~\cite{Thimm2016,FGR2022,KESHFM2022,KM2023}.

Except for special subclasses of Boolean formulas, exact minimal model counting is in $\#\co$-$\np$-complete~\cite{KK2003}.
The counting problem is often more challenging than decision problems as it requires efficiently tracing the entire search space. 
One of the key techniques that enable model counters to navigate combinatorial spaces efficiently is {\em knowledge compilation}. 
This technique involves transforming an input formula into a specific representation that facilitates model counting efficiently based on the size of the new representation~\cite{Darwiche2002,Thurley2006}. 
The knowledge compilation was applied to propositional model counting and probabilistic inference over a decade ago~\cite{CD2008}.
Since then, significant advancements have been made in the domain of model counting and its applications~\cite{SMSR2019,KJ2021}.
A natural question that arises in the literature: {\em can we design a knowledge compiler for minimal model counting?} To the best of our knowledge, no efforts have been made to extend knowledge compilation to minimal model counting.

In this work, we propose a {\em top-down} knowledge compiler, named \toolname, for minimal model counting. 
We demonstrate that, under certain structural conditions of the input Boolean formula, the counting of minimal models is in $\#\p$.
In general, \toolname~searches through an overapproximation of minimal models and extracts the minimal models using a well-known concept, called {\em justification}. 
Justification is not a new idea; it is, in fact, well-studied in {\em answer set programming}~\cite{MT1999,AGS2006}.
In this context, \toolname~shares high-level similarities with one of the recent answer set counters \sharpASP~\cite{KCM2024}.

The paper is structured as follows: \cref{section:preliminaries} provides the background necessary to understand \toolname~and related works in the area of minimal model counting.
\Cref{section:methodology} presents the proposed knowledge compiler, \toolname. \Cref{section:proof} establishes the correctness of \toolname. 
Finally, we conclude our paper with suggestions for future work in~\cref{section:conclusion}.

\section{Preliminaries}
\label{section:preliminaries}
In the paper, we use basic graph terminologies and interested readers are referred to~\cite{graphtheory}.

In propositional satisfiability, a \emph{propositional variable} $v$ takes a value from the domain $\be = \{0,1\}$, equivalently $\{\mathsf{false}, \mathsf{true}\}$. 
A \emph{literal} $\ell$ is either a variable $v$ (positive literal) or its negation $\neg{v}$ (negative literal).
A \emph{clause} $C$ is a {\em disjunction} of literals, denoted as $C = \bigvee_{i} \ell_i$. 
A Boolean formula  $F$, in \emph{Conjunctive Normal Form (CNF)}, is a {\em conjunction}
of clauses, represented as $F = \bigwedge_{j} C_j$. We use the notation $\var{F}$ to denote the set of propositional atoms or variables within  
$F$. 

An assignment $\tau$ is a mapping $\tau: X \rightarrow \{0,1\}$, where $X \subseteq \var{F}$.  For a variable $v \in X$, we define
$\tau(\neg{v}) = 1 - \tau(v)$. 
An assignment $\tau$ satifies a literal $\ell$, denoted as $\tau \models \ell$, if $\tau$ evaluates $\ell$ to \true.
Similarly, an assignment $\tau$ satifies a clause $C$, denoted as $\tau \models C$, if $\tau$ evaluates one of its literals to \true.
The assignment $\tau$ over $\var{F}$ is a {\em model} of $F$ if $\tau$ evaluates $F$ to be \true. 
We use the notation $\Sol{F}$ to denote the number of models of formula $F$.
Given a CNF formula $F$ and an assignment
$\tau: X \rightarrow \{0,1\}$, where $X \subseteq \var{F}$, 
the \textit{unit propagation} of $\tau$ on $F$, denoted
$F|_{\tau}$, is recursively defined as follows:

$F|_{\tau} = \begin{cases}
    1 & \text{if $F \equiv 1$}\\
    F'|_{\tau} & \text{if $\exists C \in F$ s.t. $F' = F \setminus \{C\}$,} \text{$\ell \models C$ 
    } \\%$\phi \equiv \phi' \wedge C$, $\ell \in C$ and $\tau(\ell) = 1$} \\
    F'|_{\tau} \cup \{C'\}  & \text{if $\exists C \in F$ s.t. $F' = F \setminus \{C\}$,} \text{$\ell \in C$,}  \text{$C' = C \setminus \{\ell\}$} 
       \text{ and $\tau \not\models \ell$}
\end{cases}$

\noindent Given a CNF formula $F$ and an assignment $\tau$, $F|_{\tau}$ {\em unit propagated} $\ell$ if $\{\ell\} \in F|_{\tau}$.

\paragraph{Minimal Model.} To define minimal models of a propositional formula $F$, we introduce an {\em ordering operator} over models~\cite{JGM2010}.
For two given models $\tau_1$ and $\tau_2$, $\tau_1$ is called {\em smaller} than $\tau_2$, denoted as $\tau_1 \leq \tau_2$, if and only if for each $x \in \var{F}$, $\tau_1(x) \leq \tau_2(x)$. 
We define $\tau_1$ as {\em strictly smaller} than $\tau_2$, denoted as $\tau_1 < \tau_2$, if $\tau_1 \leq \tau_2$ and $\tau_1 \neq \tau_2$.
A model $\tau$ is a {\em minimal model} of $F$ if and only if $\tau$ is a model of $F$ and $\not \exists \tau_2 \models F$ such that $\tau_2 < \tau$. 
We use the notation $\mmodel{F}$ to denote minimal models of $F$.
The minimal model counting problem seeks to determine $\Card{\mmodel{F}}$, denoted as $\mathsf{CntMM}(F)$.
In this paper, we sometimes represent models by listing the variables assigned to \true. For 
example, let $\var{F} = \{a,b,c\}$ and under the model $\tau = \{a\}$,  
$\tau(a) = \true$ and $\tau(b) = \tau(c) = \false$.

\paragraph{Minimal Model Checking.} Given a propositional formula $F$ and a model $\tau \models F$, checking whether $\tau$ is a minimal model of $F$ is in \conpcomplete~\cite{Cadoli1992b}. The minimal model checking subroutine checks the satisfiability of 
the formula $Q(F, \tau) = F \wedge \bigwedge_{x:\tau(x) = 0} \neg{x} \wedge \bigvee_{x:\tau(x) = 1} \neg{x}$. 
If $Q(F,\tau)$ is unsatisfiable, the model $\tau$ is a minimal model of $F$.
Otherwise, there exists a model $\tau_2 \models F$ such that $\tau_2 < \tau$, indicating that $\tau$ is not a minimal model. 
Within minimal models of $F$, each of the variables assigned to \true is {\em justified}~\cite{BH1989,KS2018}; which 
implies that if $\tau \in \mmodel{F}$ ceases to be a model of $F$ if any of its variables assigned \true are flipped to \false.

\paragraph{Dependency Graph.} Akin to {\em dependency graph} in answer set programming~\cite{KS1992}, we define a dependency graph for a given Boolean formula. 
For Boolean formula $F$, the dependency graph $\dependency{F} = (V,E)$ is a directed graph 
where node set $V = \var{F}$ and edge set $E = \{(a,b) | \{a,b\} \subseteq \var{F}, \exists C \in \clause{F} \text{ such that } \neg{a} \in C \text{ and } b \in C\}$; i.e., there is an arc in $\dependency{F}$ from node $a$ to $b$ if variables $a$ and $b$ occur in a clause $C$ as negative and positive literals, respectively. 
A Boolean formula is called {\em acyclic} (or {\em cyclic} resp.) if there is no cycle (or if there exists some cycle resp.) in $\dependency{F}$.
Additionally, a Boolean formula is {\em head-cycle free} when there does not exist any cycle in $\dependency{F}$ that contains two variables $a$ and $b$ such that $a$ and $b$ occur as positive literals in a clause $C \in \clause{F}$.

\subsection{Related Works}
\label{section:relatedwork}
To the best of our knowledge, the literature on minimal model counting is limited. 
The complexity of counting minimal models for specific subclasses of Boolean formulas, such as Horn, dual Horn, bijunctive, and affine, has been shown to be $\#\p$~\cite{DH2008}. 
For head-cycle-free CNF, the counting of minimal models is in $\#\p$~\cite{ABFL2017}.
Salhi~\cite{Salhi2019} analyzed the complexity of enumerating all minimal models of special subclasses of the Boolean formula. 
However, the general case complexity is in $\#\co$-$\np$-complete~\cite{KK2003}.

\section{Knowledge Compilation for Minimal Model Counting}
\label{section:methodology}
In this section, we present our knowledge compiler \toolname, designed for counting the number of minimal models. 
First, we introduce a formula called the {\em forced} formula, which is sufficient for counting minimal models in acyclic Boolean formulas.
For cyclic Boolean formulas, where the forced formula is insufficient for counting minimal models, we propose a specialized Boolean formula, referred to as the {\em copy formula}.
The proposed copy formula shares high-level similarities with the copy formula introduced in the answer set counter, \sharpASP~\cite{KCM2024}.

\subsection{Forced Formula}
Given a Boolean formula $F$, we will introduce a new Boolean formula $\forced{F}$, which 
ensures that if a variable $x$ is assigned \true, then this assignment is forced by one of its clauses. 
A clause $C$ forces a variable $x$ to be \true if $x$ appears in $C$ as a positive literal. 
We use the notation $\forced{F,x}$ to denote these clauses that force $x$ to be \true within formula $F$.
More specifically, if variable $x$ is assigned \true, then $\exists C \in F$ such that $x$ belongs to $C$ as a positive literal and $C \setminus \{x\} = \bot$, meaning clause $C$ is falsified if $x$ is flipped back to \false. 
For a given formula $F$, the formula $\forced{F}$ is defined as follows:
\begin{align*}
    \forced{F} = \bigwedge_{x \in \var{F}} \bigg(x \rightarrow \bigvee_{C \in \forced{F,x}} \neg(C \setminus \{x\}) \bigg)
\end{align*}
\begin{lemma}
    \label{lemma:justifynoncyclicatom}
    For acyclic Boolean formula $F$, $\cntmm{F} = \Sol{F \wedge \forced{F}}$
\end{lemma}

\begin{example}
    Consider the Boolean formula $F = (a \vee b) \wedge (b \vee c) \wedge (c \vee a)$. 
    $\forced{F} = (a \rightarrow (\neg{b} \vee \neg{c})) \wedge (b \rightarrow (\neg{a} \vee \neg{c})) \wedge (c \rightarrow (\neg{a} \vee \neg{b}))$. 
    
    The formula $F$ is acyclic and it has three minimal models. While the formula $F$ has four models but $F \wedge \forced{F}$ has three models, which corresponds to the number of minimal models of $F$. 
    Specifically, each minimal model of $F$ satifies $F \wedge \forced{F}$, whereas there is a model $\{a,b,c\} \models F$, which is not a minimal model of $F$ and does not satifies $\forced{F}$.
\end{example}

Note that $\forced{F}$ might introduce auxiliary variables. The introduction of these auxiliary variables does not affect the count of minimal models if we use {\em Tseitin encoding}~\cite{Tseitin1983} for CNF conversion.
\subsection{Copy Formula}
For cyclic formula, we introduce another Boolean formula, called {\em copy formula}, which is similar to the {\em copy operation} introduced in the context of justification within the answer set counter \sharpASP~\cite{KCM2024}.
A notable similarity between \sharpASP~and \toolname~is their introduction of {\em copy variables}. 
Moreover, the copy operation of \toolname~extends $\copyoperation{}$ operation introduced for {\em normal logic programs}~\cite{KCM2024}. 
Given a propositional formula $F$, the $\copyoperation{F}$ consists of three types of implications as follows:

\begin{enumerate}
    \item \label{l1:type1} (type $1$) for every variable $x \in \var{P}$, the implication $\copyatom{x} \rightarrow x$ is in $\copyoperation{P}$.
    \item \label{l1:type2} (type $2$) for every clause $a_1 \vee \ldots a_k \vee \neg{b_1} \vee \ldots \neg{b_{\ell}} \in \clause{F}$, where
        $\{a_1, \ldots a_k\} \neq \emptyset$, 
        the implication $\copyatom{b_1} \wedge \ldots \copyatom{b_{\ell}} \rightarrow \copyatom{a_1} \vee \ldots \copyatom{a_k}$ is in $\copyoperation{P}$.
    \item \label{l1:type3} (type $3$) for every variable $x \in \var{F}$ where $\not \exists C \in \clause{F}, x \in C$, the implication $x \rightarrow \false$.
\end{enumerate}
Note that  if a clause contains no positive literals, a type~\ref{l1:type2} implication is not introduced for that clause. Additionally, if a variable does not occur as a positive literal in any clause, the variable must be \false in each minimal model. Consequently, a type~\ref{l1:type3} implication is introduced for such variables.
\begin{example}
    Consider Boolean formula $F = (\neg{a} \vee b) \wedge (\neg{b} \vee c) \wedge (\neg{c} \vee a)$.
    
    \noindent The Boolean formula is cyclic because there is a cycle between atoms $a,b$ and $c$.
    For formula $F$, $\forced{F} = (a \rightarrow c) \wedge (b \rightarrow a) \wedge (c \rightarrow b)$ and  
    $\copyoperation{P} = (\copyatom{a} \rightarrow \copyatom{b}) \wedge (\copyatom{b} \rightarrow \copyatom{c}) \wedge (\copyatom{c} \rightarrow \copyatom{a}) \wedge (\copyatom{a} \rightarrow a) \wedge (\copyatom{b} \rightarrow b) \wedge (\copyatom{c} \rightarrow c)$.
    The formula $F$ has a single minimal model $\{\}$, while two assignments satisfy $F \wedge \forced{F}$. The assignment $\{a,b,c\}$ satifies $F \wedge \forced{F}$, but it is not a minimal model of $F$.
\end{example}

\subsection{Knowledge Compiler: \toolname}
Similar to \sharpASP, we introduce the knowledge compiler \toolname, which operates on a tuple of Boolean formulas $(P,G)$. 
In the context of minimal model counting, given a Boolean formula $F$, $P = F \wedge \forced{F}$ and $Q = \copyoperation{F}$. 
In the tuple representation $(P,Q)$, the formula $P$ overapproximates the search space of minimal models of $F$, while $Q$ extracts the minimal models of $F$ from $P$ through justification. 
Given a propositional formula $F$, we introduce the key components of knowledge compilation on $(P,G)$. 

\paragraph{Decomposition.}
Model counters often \textit{decompose} the input CNF formula into \textit{disjoint subformulas} 
to enhance computational efficiency~\cite{Thurley2006}. Specifically, for two formulas $\Phi_1$ and $\Phi_2$, if $\var{\Phi_1} \cap \var{\Phi_2} = \emptyset$, 
then $\Phi_1$ and $\Phi_2$ are \textit{decomposable}. This means that we can count the number of models of $\Phi_1$ and $\Phi_2$ separately and multiply 
these two counts to get the number of models of $\Phi_1 \wedge \Phi_2$. 
We define \emph{component decomposition} over the tuple representation $(P, Q)$ as follows:
\begin{definition}
\label{def:decomposition}
    The tuple $(P_1 \wedge P_2, Q_1 \wedge Q_2)$ is decomposable to $(P_1,Q_1)$ and $(P_2,Q_2)$ if and only if $(\var{P_1} \cup \var{Q_1}) \cap (\var{P_2} \cup \var{Q_2}) = \emptyset$.
\end{definition}
Finally, \Cref{lemma:establishdecomposition} provides a guarantee regarding the correctness of our proposed definition of decomposition in counting the number of minimal models.
\begin{lemma}
    \label{lemma:establishdecomposition}
    Let $(P_1 \wedge \ldots P_k,  Q_1 \wedge \ldots Q_k)$ is decomposed to $(P_1, Q_1),  \ldots, (P_k, Q_k)$, then 
    $\cntmm{P_1 \wedge \ldots P_k,Q_1 \wedge \ldots Q_k} = \cntmm{P_1,Q_1} \times \ldots \cntmm{P_k,Q_k}$
\end{lemma}

\paragraph{Determinism.}
Model counters utilize the concept of \textit{determinism}~\cite{Darwiche2002}, which involves splitting the formula by assigning one of its variables to either \false or \true.
The number of models of $\Phi$ is then determined by simply summing the number of models in which a variable $x \in \var{\Phi}$ is assigned to \false and \true.  
This approach can similarly be applied to minimal model counting using our pair representation.
\begin{lemma}
    \label{lemma:determinismequation}
    Given a Boolean formula $F$,
      \begin{align}
          \cntmm{P,Q} &= \cntmm{P|_{\neg{x}},Q|_{\neg{x}}} + \cntmm{P|_{x},Q|_{x}},\nonumber \\ &~~ \text{for all } x \in \var{P} \label{eq:determinism} \\% \nonumber \\
          \cntmm{\bot,Q} &= 0 \nonumber\\ 
          \cntmm{\emptyset,Q} &= 
          \begin{cases}
              1 & \text{if $Q = \emptyset$ or $\forall \copyatom{x} \in \var{Q}, x$ is justified} \label{eq:basecase}\\
              0 & \text{otherwise}
          \end{cases}
      \end{align}
\end{lemma} 
\paragraph{Checking Justification: base case.}
When no variables of $P$ remain to be assigned, it is necessary to verify whether each assigned variable is justified. To justify each atom, a SAT solver must be invoked over the Boolean formula 
$Q$. Variables assigned to \false are trivially justified. Therefore, if a variable 
$x$ is assigned to \false, then the copy variable $\copyatom{x}$ is unit propagated to \false due to the type~\ref{l1:type1} implication. If 
$\exists \copyatom{x} \in \var{Q}$, then the corresponding variable $x$ is assigned to \true within the subformula $P$.

The justification checking subroutine verifies the justification for each atom present in $\copyatom{x} \in \var{Q}$ and checks the satisfiability of the formula 
$Q \wedge \bigvee_{\copyatom{x} \in \var{Q}} \neg{\copyatom{x}}$. If 
$Q \wedge \bigvee_{\copyatom{x} \in \var{Q}} \neg{\copyatom{x}}$ is unsatisfiable, then each atom 
$x$ such that $\copyatom{x} \in \var{Q}$ is justified under the partial assignment. Otherwise, 
$\exists \copyatom{x} \in \var{Q}$ such that $x$ is not justified.

\begin{proposition}
    Given a Boolean formula $F$, the number of minimal models can be determined by~\Cref{lemma:determinismequation}: $\cntmm{F} = \cntmm{P,Q}$, where $P = F \wedge \forced{F}$ and $Q = \copyoperation{F}$.
\end{proposition}
\addtocounter{example}{-1}
\begin{example}[continued]
    Consider Boolean formula $F = (\neg{a} \vee b) \wedge (\neg{b} \vee c) \wedge (\neg{c} \vee a)$. For $F$, $\tau_1 = \{\} \in \mmodel{F}$ because $\copyoperation{F}|_{\tau_1} = \emptyset$. While $\tau_2 = \{a,b,c\} \not \in \mmodel{F}$ because $\copyoperation{F}|_{\tau_2} \wedge (\neg{\copyatom{a}} \vee \neg{\copyatom{b}} \vee \neg{\copyatom{c}})$ is satifiable. 
\end{example}

\section{Theoretical Correctness of \toolname}
\label{section:proof}
In this section, we establish the theoretical correctness of \toolname. 
\paragraph{Proof of~\Cref{lemma:justifynoncyclicatom}}
\begin{proof}
    We use proof by contradiction. Assume that 
    there is a model $\tau \models F \wedge \forced{F}$ and $\tau$ is not a minimal model of $F$.
    Thus, under $\tau$, there exists an atom $x_1$, which is not justified. More specifically, 
    there exists a clause $C_1 \in \clause{F}$ that forces $x_1$ to be \true under $\tau$.
    When $x_1$ is not justified, there are two possible cases:
    \begin{itemize}
        \item there is a variable $v_+$ that appears in $C_1$ as a positive literal such that $\tau(v_+) = \false$ and $v_+$ must be assigned to \true to satisfy $C_1$. More specifically, the clause $C_1$ forces the variable $v_+$. However, this case cannot evidence about the lack of justification of $x_1$  
        because if $v_+$ is assigned to \false, then $x_1$ must be assigned to \true to satisfy the clause $C$. So, it contradicts that variable $x_1$ is not justified. 
        \item there is a variable $v_-$ that appears in $C_1$ as a negative literal such that $\tau(v_-) = \true$ and $v_-$ must be assigned to \false to satisfy $C_1$. Note that $\tau(v_-) = \true$, which implies that 
        variable $v_-$ itself is not justified under $\tau$. Let denote $v_-$ as $x_2$ and there is another clause $C_2$ that forced $x_2$ to be true under assignment $\tau$.
        As a result, there is a sequence of variable $\{x_i\}$, starting from $i = 1$. Note that neither of $x_i$ is on cycles. 
        So, there is an unbounded number of atoms that are not justified, which is a contradiction. 
    \end{itemize}
\end{proof}
\paragraph{Proof of~\Cref{lemma:establishdecomposition}}
\begin{proof}
    Within the tuple representation of $(P,Q)$, the formula $P$ provides 
    an overapproximation of minimal models. Let the formula $P$ be written as 
    $P_1 \wedge \ldots P_k$, where $\forall_{i \neq j} \var{P_i} \cap \var{P_j} = \emptyset$ and $i,j \in [1,k]$.
    It follows that each model $\tau \models P$ can be written as 
    $\tau = \tau_1 \cup \ldots \tau_k$, where $\forall_{i} \var{P_i} = \var{\tau_i}$ and vice versa. 
    Thus, decomposing $P$ into $P_1, \ldots, P_k$ does not change the overapproximation of the minimal models of $F$. Concurrently, the formula $Q$ is used to check justification of atoms. 
    Assume that the formula $Q$ be written as $Q = Q_1 \wedge \ldots Q_k$.
    Assume a copy variable $\copyatom{x} \in Q_i$, the justification of $\copyatom{x}$ does not depend on the variables $\var{Q} \setminus \var{Q_i}$. 
    Thus, decomposing $Q$ into disjoint components does change justification of atoms.
    As a result, decomposing the tuple representation $(P,Q)$ into $(P_1, Q_1), \ldots, (P_k, Q_k)$ preserves 
    the count of the number of minimal models.
\end{proof}
\paragraph{Proof of~\Cref{lemma:determinismequation}}
\begin{proof}
    The models of formula $P$ are closed under unit propagation. 
    \toolname~assigns only non copy variables and checks justifications over copy variables. Thus, the unit propagation by non copy variables preserves the minimal models of the input formula.
    In fact, the determinism (\Cref{eq:determinism}) simply partitions the minimal models of $F$ based on {\em decision variables} $x$.

    Within the formula $Q$, a copy variable $\copyatom{x}$ is unit propagated to either $0$ or $1$, when the assignment to $x$ is justified.
    At the base case, when $P = \emptyset$, the presence of a copy variable $\copyatom{x} \in \var{Q}$ indicates that the justification of $x$ is not proven yet 
    and \toolname~needs to check its justifications.
    In the base case (\Cref{eq:basecase}), $Q= \emptyset$ indicates that the trivial case, where no variable is left to check justifications. Similarly, if $Q \wedge \bigvee_{\copyatom{x} \in \var{Q}} \neg{\copyatom{x}}$ is unsatifiable, then it implies that each atom is justified. 
    In these above cases, \toolname~returns $1$ denoting that the partial assignment can be extended to a minimal model. 
    In contrary, when $Q \wedge \bigvee_{\copyatom{x} \in \var{Q}}$ is satisifiable, at least one of the atoms can be set to \false indicating the lack of justification of at least one atom; thus, \toolname~returns $0$ in the case. 
\end{proof}

\section{Conclusion}
\label{section:conclusion}
In this work, we introduce \toolname, the first knowledge compiler designed for minimal model counting. 
For acyclic formulas, \toolname~reduces minimal model counting to $\#\p$. 
For cyclic formulas, \toolname~employs some sophisticated techniques based on checking justifications. 
One of the key scalability issues of \toolname~is that it invokes an SAT oracle at each base case. 
One key scalability issue of \toolname~is its reliance on invoking a SAT oracle at each base case. 
Future research will focus on implementing the knowledge compiler, improving its scalability on larger benchmarks, and exploring real-world applications of minimal model counting.

\nocite{*}
\bibliographystyle{eptcs}
\bibliography{example}
\end{document}